\documentclass[12pt]{article}
\usepackage{latexsym}  
\def\a{\alpha} 
\def\b{\beta}
\def\g{\gamma}

\def\d{\delta}

\def\l{\lambda}

\def\m{\mu} 
\def\n{\nu}   
\def\h{\eta}
\def\f{\phi}   
\def\r{\rho}   
\def\o{\omega}
 
\def\s{\sigma}

\def\x{\xi}
   
\def\e{\epsilon}
\def\ve{\varepsilon}

\def\pa{\partial}

\def\be{\begin{equation}}   
\def\ee{\end{equation}}   
\def\bqn{\begin{eqnarray}}   
\def\eqn{\end{eqnarray}}
\def\se{\stackrel{e}}
\def\su{\stackrel{u}}
\def\sv{\stackrel{v}}

\def\ca{{\cal A}}

\def\cd{{\cal D}}

\def\cl{{\cal L}}

\begin{document}
%
%%%%%%%%%%%%%%%%%%%%%%%%%%%%%%%%%%%%%%%%%%%%%%%%%%%%%%%%%%%%%%%%%%
\addtolength{\topmargin}{-2cm}
\addtolength{\textheight}{3.5cm}
%\addtolength{\oddsidemargin}{-1cm}
\addtolength{\textwidth}{1.5cm}
\addtolength{\footskip}{1cm}
\sloppy
%%%%%%%%%%%%%%%%%%%%%%%%%%%%%%%%%%%%%%%%%%%%%%%%%%%%%%%%%%%%%%%%%%% 

\begin{titlepage}
\begin{flushright}
ULB-TH-01/41\\
\end{flushright}
\begin{centering}

\vspace{.1cm}

{\huge {\bf Conformal (super)gravities with several gravitons}} 

\vspace{1.1cm}
 
{\Large Nicolas Boulanger$^{a,}$
%\footnote{"Chercheur F.R.I.A.", Belgium},
Marc Henneaux$^{a,b}$ and Peter van Nieuwenhuizen$^{c}$}\\
\vspace{1.5cm}

$^a$ {\small\slshape Physique Th\'eorique et Math\'ematique, 
 Universit\'e Libre
de Bruxelles,  C.P. 231, B-1050, Bruxelles, Belgium}      \\
\vspace{.1cm}
$^b$ {\small\slshape Centro de Estudios Cient\'{\i}ficos, 
Casilla 1469, Valdivia, Chile}  \\
\vspace{.1cm}
$^c$ 
{\small\slshape
 C.N.~Yang Institute for Theoretical Physics \\
 State University of New York at Stony Brook \\
 Stony Brook, NY 11794-3840, USA} \\
\vspace{.3cm} 
{\small{\tt nboulang@ulb.ac.be,henneaux@ulb.ac.be, 
vannieu@insti.physics.sunysb.edu}}
 
\end{centering}

\vspace{1.5cm}

\begin{abstract}
We construct consistent interacting gauge theories for $M$
conformal massless
spin-2 fields ("Weyl gravitons") with the
following properties: (i) in the free limit, each
field fulfills the equation ${\cal B}^{\mu \nu} = 0$, where
${\cal B}^{\mu \nu}$ is the linearized Bach tensor,  
(ii) the interactions contain no more than four derivatives,
just as the free action and (iii) the internal metric for the Weyl gravitons
is not positive definite.  
The interacting theories are obtained by gauging
appropriate non-semi-simple extensions of the conformal algebra $so(4,2)$ 
with  commutative,
associative algebras of dimension $M$. By writing the action in terms of 
squares of supercurvatures, 
supersymmetrization is immediate and leads to consistent
conformal supergravities with $M$ interacting gravitons.
\end{abstract}

\vfill
\end{titlepage}    

\section{Introduction}
Recently, no-go theorems have been proven that
prevent interactions between a set of massless symmetric
fields $h^a_{\mu \nu}$  ($a = 1, \cdots, M$) described in
the free limit either
by the Pauli-Fierz \cite{BDGH} or the linearized Weyl \cite{Boulanger:2001he}
Lagrangian.  These theorems were obtained by using the
antifield-based cohomological approach to the
problem of consistent interactions \cite{Barnich:vg,MH1}.

A crucial assumption underlying the no-go theorems 
is that the metric in the internal space of
the fields $h^a_{\mu \nu}$ be positive definite.
This assumption appears to be quite reasonable in the Pauli-Fierz
case, where it guarantees that the free (classical) theory has non-negative
energy, leading to a quantum theory free of negative norm states,
or "ghosts".
It is less motivated, however, 
in the conformal case where the classical
energy is unbounded from below even for a positive definite internal
metric.  For this reason, it is of interest to investigate
theories in which positive definiteness of the internal metric
is relaxed. 

If the assumption on the metric in internal space is dropped, 
interactions become possible.  An explicit example involving
two (Weyl) "gravitational" fields $h^1_{\a \b}$
and $h^2_{\a \b}$ in interaction was given
in \cite{Boulanger:2001he}. The action reads
\be
S[h^a_{\m \n}] = \frac{1}{\alpha} \int d^4 x \sqrt{-g} \, h^2_{\m \n} \, 
B^{\m \n},
\label{strange0}
\ee
where
\be
B_{\m \n}=2\nabla^{\a}\nabla^{\r}C_{\r\m\n\a}+R^{\l\r}C_{\l\m\n\r}
\ee
is the Bach tensor of the metric
$g_{\a \b} = \eta_{\a \b} + \a h^1_{\a \b}$ and $g$ its
determinant.    The parameter $\alpha$ is a deformation parameter.
In the  limit $\alpha \rightarrow 0$, one recovers the free
theory, with Lagrangian $\sim    h^2_{\m \n} D^{\m \n \a \b} h^1_{\a \b}$
where $D^{\m \n \a \b}$ is the fourth order differential operator
appearing in the (linearized)  equations of motion.  The
free equations of motion read ${\cal B}^{a \, \m \n} = 0$,
where ${\cal B}^{a \, \m \n}$ is the linearized Bach tensor, 
${\cal B}^{a \, \m \n} = D^{\m \n \a \b} h^a_{\a \b}$ ($a=1,2$)
(see \cite{Boulanger:2001he} for more information).             
The gauge transformations are 
\begin{eqnarray}
%\frac{1}{\alpha} \d_{\h^a,\f^a}g_{\m\n}
%&=&\h^1_{\m;\n}+\h^1_{\n;\m}+2\f^1 g_{\m\n} ,
\d_{\h^a,\f^a}h^1_{\m\n}
&=&\h^1_{\m;\n}+\h^1_{\n;\m}+2\f^1 g_{\m\n} ,
\label{strange1}\\
\d_{\h^a,\f^a} h^2_{\m \n} &=& \a {\cal L}_{\h^1} h^2_{\m \n}
+ 2 \a \f^1 h^2_{\m \n} + \h^2_{\m;\n}+\h^2_{\n;\m}
+2\f^2 g_{\m\n},
\label{strange2}
\end{eqnarray}
where covariant derivatives ($;$) are computed with the
metric $g_{\a \b}$ and where ${\cal L}$ denotes the Lie derivative
${\cal L}_{\h}h_{\m\n}=\pa_{\m}\h^{\r}h_{\r\n}+\pa_{\n}\h^{\r}h_{\r\n}
-\h^{\r}\pa_{\r}h_{\m\n}$.
As one sees from the free action, the metric
$k_{ab}$ in internal space is given by $k_{12} = k_{21} = 1$,
$k_{11} = k_{22} = 0$ and so has signature $(- +)$.  By introducing
the linear combinations $\chi_{\a \b} \sim h^1_{\a \b} +
h^2_{\a \b}$ and
$\psi_{\a \b} \sim h^1_{\a \b} - h^2_{\a \b}$, one can diagonalize
$k_{ab}$ and get as free Lagrangian ${\cal L}^F \sim
\chi_{\m \n} D^{\m \n \a \b} \chi_{\a \b} - \psi_{\m \n}
D^{\m \n \a \b} \psi_{\a \b}$.

The purpose of this note is 
to provide a group-theoretical understanding of (\ref{strange0})
in terms of the conformal group along the lines of 
\cite{KTvN1,Mansouri:pr}.  We show that the above "2-Weyl-graviton
theory" (respectively, the "$M$-Weyl-graviton theory"\footnote{We shall
loosely call hereafter "$M$-Weyl-graviton theories"
theories with a set of $M$ fields
$h^a_{\m \n}$ obeying the Bach equations in the free limit.}) can
be viewed as the gauge theory of the tensor product $so(4,2) 
\otimes \ca$ of the conformal algebra $so(4,2)$ with an
irreducible, associative and commutative algebra $\ca$ of
dimension 2 (respectively, $M$).  The gauging procedure
follows the general pattern explained in \cite{Mansouri:1977ej,PvN}.

The advantage of the group-theoretical viewpoint is that it
allows for a direct supersymmetrization.  
We restrict ourselves to $N=1$ (simple) conformal supergravities, 
{\it{i.e.}}, supergravities with one graviton. These we extend to theories
with $M$ gravitons (whereas $N$-extended supergravities still always have
one graviton).
One must simply
repeat the steps worked out in \cite{PvN,KTvN2,TvN} for 
the one-multiplet case.  We provide a general existence
proof of 
the $M$-multiplet conformal supergravities and give a
constructive method for deriving the Lagrangian from
the 1-multiplet theory.

At the quantum level, conformal and superconformal theories are, according
to the usual perturbative approaches, not unitary. The counting of states is
more subtle \cite{Lee:cp}. These drawbacks are 
of course also present in their multi-multiplet "daughters".
For this reason, most physicists have used
the conformal supergravities as building blocks and backbones of ordinary
supergravities without attributing much direct physical meaning to them.
However, another point of view, which is increasingly adopted, is to view
conformal (super)gravities as effective field theories, in which case 
standard issues
of unitarity would no longer apply \cite{TseytlinSUGRA@25}.
From this point of
view, conformal supergravities may become of great
importance in the future.
We should also mention that unconventional possibilities for actually making 
sense out of
theories with ghosts have been explored recently
in terms of appropriate averages over the ghost states \cite{Hawking:2001yt},
or through dualities \cite{Hull:1998vg,Hull:2001ii}
 (for some earlier analysis, see also 
\cite{Tseytlin:1995yw}). 

Although we concentrate on conformal gravity
for reasons explained above, similar considerations
apply to standard gravity, provided one replaces the conformal
group by the (anti) de Sitter group or its 
Poincar\'e contraction \cite{MacDowell:1977jt}.  In that case
also, one can construct interacting multi-graviton theories
if the metric in internal space is not positive definite
\cite{Wald1,BDGH}.  An explicit example,
analogous to (\ref{strange0}), was given in \cite{ovrut}.
The group-theoretical approach
provides considerable insight into the algebraic structures underlying
the interactions discussed
by Wald \cite{Wald2} by relating them to ordinary
Lie algebras and their gaugings. It also yields
straightforwardly real-valued (as opposed to
algebra-valued) actions.  Finally, the approach 
significantly clarifies the
construction of the supersymmetric actions studied in
\cite{Anco:1998mf}.

%%%%%%%%%%%%%%%%%%%%%%
\section{Algebraic preliminaries}
\label{section2}
%%%%%%%%%%%%%%%%%%%%%%%%%%%%%%%%%%%%%%%%%%%%%%%%%%%%%%%%%%%%%%%%%%%%%%%%%%
\setcounter{equation}{0}  
\setcounter{theorem}{0}  
\setcounter{lemma}{0}  
%~~~~~~~~~~~~~~~~~~~~~~~
\subsection{Associative, commutative algebras}
We first recall the results of \cite{BDGH,Boulanger:2001he}, obtained by
using the cohomological tools developed in \cite{DV0,locality,DV,BBH1,BBH2}
for dealing with the local cohomologies $H(s\vert d)$ ($s$ being
the BRST differential and $d$ the spacetime exterior derivative)
relevant to the
gauge-consistent deformations of a gauge-invariant theory
(by gauge-consistent we mean that a $M$-graviton theory will have $M$ times 
the number of gauge symmetries of the corresponding one-graviton theory).
The various fields $h_{\m\n}^a$, $a=1,\ldots M$ of the
given set of (Weyl) gravitons are assumed to live
in an internal $M$-dimensional algebra $\ca$, i.e.,
a (real) vector space endowed with a bilinear map
$\star : \ca \times \ca \rightarrow \ca$, {\it{i.e.}} a tensor $a^a_{~bc}$ 
of type $(1,2)$ over $\ca$, generally referred to as "multiplication".
Upon expansion into a basis
$\left\{ E_a \right\}_{a=1}^M$ 
\be
v=v^a E_a, ~~~u=u^a E_a,
\ee 
the product reads
\be
u \star v=z=z^aE_a=u^b a_{~bc}^a v^c E_a.
\ee    
The $a^a_{~bc}$ define the interactions \cite{BDGH,Boulanger:2001he}.  
The algebra $\ca$
is also equipped with an internal scalar product, $<u,v> =
k_{ab} u^a v^b$, defined by the quadratic part of the
action (i.e., the free action).

The tensor $ a^a_{~bc} $ and the metric $k_{ab}$ were shown to be 
constrained by the following identities for the deformation to be
consistent :
\bqn
a^a_{~bc}&=&a^a_{~cb}
\label{commu} \\
a^a_{~b[c}a^b_{~d]e}&=&0
\label{asso} \\
k_{ad}a^d_{~bc}\equiv a_{abc}&=&a_{bac}
\label{sym} 
\eqn
The first equation (\ref{commu}) implies that the algebra $\ca$ is commutative,
 (\ref{asso}) that it is associative, and (\ref{sym}) that the metric is 
an invariant tensor\footnote{the conditions (\ref{commu}) and (\ref{sym})
together imply that $a_{abc}=a_{(abc)}$, so we will sometimes use the 
term "symmetric" when referring to the algebras endowed with such a $a_{abc}$ 
tensor.},
\bqn
u \star v &=&  v \star u 
\label{commu"} \\
(u \star v) \star w &=& u \star (v \star w)
\label{asso"} \\
<u \star v, w> &=& <u, v\star w>.
\label{sym"}
\eqn       
The same conditions arise in the Einstein case.  The conditions
(\ref{commu}) and (\ref{asso}) 
were first derived in that context in the
pioneering work \cite{Wald1}, where the emphasis was put
on the consistency of the gauge transformations and their algebra.
The further symmetry condition (\ref{sym"}) emerges from  
the analysis of the action itself.
Its importance was particularly stressed
in \cite{BDGH,Boulanger:2001he}. 

The structure constants $ a^a_{~bc} $ characterize actually
the only possible consistent deformations of the free action
under the condition of Lorentz invariance and the restriction that
the complete action should contain no more derivatives (namely 4) than
the free action \cite{Boulanger:2001he}. The conditions (\ref{commu"}), 
(\ref{asso"})
and (\ref{sym"}) guarantee consistency of the deformation up to
second order included.   It follows from the existence of actions
explicitly given below that consistency holds in fact to all orders.

Taking the internal metric $k_{ab} $ 
to be positive-definite ($k_{ab} =\d_{ab}$), i.e., starting
with a free action that is the sum (with only plus signs)
of linearized Weyl (or Pauli-Fierz) actions, the
algebra $\ca$ can be shown to be the direct sum  of one-dimensional
ideals, which implies the existence of
a basis where $a^a_{~bc} =0$ whenever two indices are different 
\cite{BDGH} :
the cross-interactions between the
various Weyl-gravitons can be removed by redefinitions.
If $k_{ab}$ is of mixed signature, however,
the algebra $\cal A$ need not be trivial, and
one can construct truly interacting multi-gravitons theories.

As shown in \cite{Wald2}, irreducible, commutative, 
associative algebras of finite dimension
can be divided into three types: 
\begin{enumerate} 
\item ${\cal A}$ contains no identity element and 
every element of ${\cal A}$ is nilpotent ($v^m = 0$ for 
some $m$). 
\item ${\cal A}$ contains one (and only one) identity element  
$e$ and no element $j$ such that $j^2 = - e$. In that case, 
${\cal A}$ contains a $(M-1)$-dimensional ideal of nilpotent elements 
and one may choose a basis $\{e, v_k \}$ ($k = 1, \cdots, M-1$) 
such that all $v_k$'s are nilpotent. 
\item ${\cal A}$ contains one identity element 
$e$ and an element $j$ such that $j^2 = -e$.  The algebra ${\cal A}$  
is then of even dimension $ M = 2 m$, and there exists a 
$(2(m-1))$-dimensional ideal of nilpotent elements.  One can 
choose a basis $\{e, v_k, j, j \cdot v_k \}$ ($k = 1, \cdots, m-1$) 
such that all $v_k$'s are nilpotent. 
\end{enumerate} 

Algebras which are commutative, associative, with a unity and  a 
nondegenerate inner product which makes the algebra symmetric 
({\it{i.e.}} a nondegenerate {\it{invariant}} inner product )
are called Fr\"obenius algebras\footnote{We thank Giulio Bonelli for having
pointed this out to us}.

%**************************************
\subsection{Tensor product of a Lie algebra and an associative, commutative
algebra}
%**************************************
We have just seen that multi-graviton theories require associative,
commutative algebras.  On the other hand, the theory of a
single type of Weyl gravitons requires the Lie algebra
of the conformal group \cite{KTvN1}.  How can the two be put
together?

Let $L$ be a Lie algebra (e.g., the Lie algebra of the
conformal group) 
whose generators we denote by $T_A$. One has
\be
[T_A,T_B]=f_{AB}^{~~C}T_C
\ee
where the $f_{AB}^{~~C}$ are the structure constants.
Consider the direct product $L'=L\bigotimes \ca$ of the Lie algebra $L$ 
with
the associative commutative symmetric $M$-dimensional algebra $\ca$.
It is easy to see that $L'$ has a natural Lie algebra structure.
An arbitrary element of $L'$ is a sum of terms of the form 
$X\otimes x$, where $X=X^AT_A \in L$ and
$x=x^aE_a\in\ca$.                    
The Lie bracket in $L'$ is defined by
\bqn
 {[.\; , .]}_{L'} : L'\times L' &\rightarrow& L'
\nonumber \\
{[{X\otimes x},{Y\otimes y}]}_{L'}  &\equiv& {[X,Y]}_{L}
\otimes x\star y.
\eqn
The bilinearity of ${[.\;, .]}_{L'}$ follows from the bilinearity
of  ${[.\;, .]}_{L}$ and $\star$.
The $L'$-bracket
is antisymmetric because
the $L$-bracket is antisymmetric and the $\star$-product is symmetric.
Similarly, the Jacobi identity holds in $L'$ because it
holds in $L$ and because $\ca$ is commutative and associative.
In general, the algebra $L'$ is not semi-simple, even if $L$ is, because
$\ca$ contains nilpotent elements\footnote{For $L'$ to be
a Lie algebra, $\ca$ need not be finite-dimensional.  In fact, the
same construction forms the basis of "loop algebras",
where $L$ is taken to be a finite-dimensional Lie algebra
and $\ca$ the (infinite-dimensional) algebra of functions on the circle. 
 Setting
$T_A^n \equiv T_A \exp(2 \pi n i \varphi)$, one
gets
$$
[T_A^m, T_A^n] = f_{AB}^{~~C} T_C^{n+m}, \; \; \; \; n,m \in Z.
$$}.

If $L$ is equipped with a quadratic form, $(X,Y) = g_{AB} X^A Y^B$,
one can extend it to $L'$ through the formula
\be
({X\otimes x},{Y\otimes y})_{L'} \equiv (X,Y)<x,y>.
\label{scalprodinprod}
\ee
Because of (\ref{sym"}), $(,)_{L'}$ is invariant in $L'$,
\be
([X\otimes x,Y\otimes y]_{L'}, Z\otimes z)_{L'}
= (X\otimes x, [Y\otimes y,Z\otimes z]_{L'})_{L'} 
\ee
whenever $(,)_{L}$ is invariant in $L$, $([X,Y],Z) = (X,[Y,Z])$.

\subsection{Examples}

{}From now on, $L$ is the conformal algebra $so(4,2)$
\bqn
&&[M_{ab},M_{cd}]=\eta_{bc}M_{ad} + \eta_{ad}M_{bc} 
- \eta_{ac}M_{bd} -\eta_{bd}M_{ac},
\nonumber \\
&&[P_a,M_{bc}]=\h_{ab}P_{c} - \h_{ac} P_d,
~~~[K_a,M_{bc}]=\h_{ab}K_{c} - \h_{ac}K_d,
\nonumber \\
&&[P_a,K_b]=2(\h_{ab} D -  M_{ab} ),
\nonumber \\
&&[P_a , D ]=  P_a, ~~~[K_a , D ] = - K_a.
\label{algebra}
\eqn                
with
bilinear form
\be
(X,Y) =\frac{1}{2} \ve_{abcd}X^{ab}
Y^{cd}
\label{scalprodconf}
\ee
considered in \cite{KTvN1} in order to construct the action.
Here, $X^{ab}$ and $Y^{ab}$ are the components of $X$ and $Y$
along the Lorentz generators.  We follow the conventions of
\cite{KTvN1,KTvN2}.  The scalar product (\ref{scalprodconf})
is not invariant under the full conformal algebra, although
it is invariant under the Lorentz subalgebra. 

When the associative algebra $\ca$ contains an identity element $e$,
the Lie algebra $L'$ contains a subalgebra isomorphic to
the conformal algebra, namely, $so(4,2) \otimes \{e \}$. 
As we shall see, the gauging leads then to Riemannian geometry
(+ extra fields).
We shall restrict the present discussion to those cases.
It is clear that case 3 in the above classification
of associative algebras is the complexification of
case 2, so we just consider this latter case.
Theories of complex gravity, {\it{i.e.}} theories with a complex
metric have been extensively studied in the past. A more recent study is in
 \cite{Wald2}.

The simplest non-trivial algebra $\ca$ is
spanned by two elements, a unit one $e$ and a nilpotent
one of order two $n$ ($e^2 = e$, $en=ne=n$, $n^2 = 0$).
The most general (invertible) internal metric $k_{ab}$
compatible with (\ref{sym}) is, up to a multiplicative
factor, given by 
\begin{equation}
\hspace{1.5cm} k_{ab} = \left( \matrix{k &1 \cr
1 & 0 \cr
} \right).
\label{PSinB}
\end{equation}
where $k$ is an arbitrary number.
Explicitly,
\bqn
&&k(e,e)=k \, ;  \; \; k(n,n)=0;
\nonumber \\
&&k(e,n)=k(n,e)=1.
\label{PSinA}
\eqn

We shall adopt the following notation :
the elements $T_A\otimes e$ of $L'$ are written as $T_A^e$, 
while the $T_A\otimes n$ are written as $T_A^n$. 
Since $L \equiv so(4,2)$ is the 15-parameter conformal algebra,
the Lie algebra $L'$ has dimension 30.
In fact, since $L$ is isomorphic to the
subalgebra $L \otimes \{e\}$, we will simply write ${T_A}$ instead of 
${T_A} {\otimes} e$.  Then the algebraic structure of $L'$ is given by 
\bqn
&&[T_A,T_B]=f_{AB}^{~~C}T_C
\nonumber \\
&&[T_A,T^n_B]=f^{~~C}_{AB}{T}^n_{C}
\nonumber \\
&&[T^n_{A},T^n_{B}]=0,
\label{bigalgebra}
\eqn
with $f^{~~C}_{AB}$ given in (\ref{algebra}).
The subspace generated by the $T^n_{A}$ is an abelian ideal.

According to (\ref{scalprodinprod}), the
scalar product (\ref{scalprodconf}) extends to $L'$ as
\be
(X,Y) = \frac{1}{2} \ve_{abcd} (X^{ab} Y^{n \; cd}
+ X^{n \; ab} Y^{cd} + k X^{ab} Y^{cd})
\label{scalprodexten2}
\ee
(using (\ref{PSinA}) and an obvious notations). 
It is invariant under the subalgebra ${\cal L} \otimes \ca$,
where ${\cal L}$ is the Lorentz subalgebra of $so(4,2)$.

The next simplest case is three-dimensional.  There are actually
two three-dimensional algebras.
The first one is 
generated by the identity and two nilpotent elements
of order 2, $n^2 = 0$, $n'^2 = 0$, such that $nn'=0$. The internal scalar
product is however degenerate. [The scalar products $<n,n>$,
$<n',n'>$ and $<n,n'>$ vanish because $n^2 = 0$, $n'^2 = 0$
and $nn'=0$ (e.g., $<n,n'>=
<n,n'e>=<nn',e>$ (symmetry) $= 0$) so there is a linear combination
of $n$ and $n'$ that is orthogonal to all other elements.]

{}For this reason, it is the other three-dimensional
irreducible, commutative, associative
algebra, which is  of most interest.  This
algebra is generated by an identity $e$ and a nilpotent
element $u$ of order 3, $u^3 = 0$.  The algebra is
spanned by the three basis elements $\left\{ e,u,v = u^2\right\}$
and the multiplication table reads
\bqn
\begin{array}{r||l|c|c|c|c}
\star & e & u & v
\\ \hline\hline
\rule{0em}{3ex}
e & e & u & v
\\ \hline
\rule{0em}{3ex}
u & u & v & 0
\\ \hline
\rule{0em}{3ex}
v & v & 0 & 0
\\ \hline
\end{array}
\eqn
The Lie algebra formed by the tensorial product of $L$ and $\ca$ possesses,
schematically, the following commutation rules :
\bqn
&&[\se{T},\se{T}]=\se{T},~~[\se{T},\su{T}]=\su{T},
\nonumber \\
&&[\se{T},\sv{T}]=\sv{T},~~[\su{T},\su{T}]=\sv{T},
\nonumber \\
&&[\su{T},\sv{T}]=[\sv{T},\sv{T}]=0
\label{2.20}
\eqn
where $[\se{T},\se{T}]=\se{T}$
really stands for $[\se{T_B},\se{T_C}]=f_{BC}^{~~A}\se{T_A}$, etc.

The (non-positive definite) internal metric $k_{ab}$ (where
$a,b$ run over
$e,u,v $) that makes the algebra symmetric is,
up to an overall multiplicative factor, 
\begin{equation}
\hspace{1.5cm} k_{ab} = \left( \matrix{k &l &1 \cr
l & 1 & 0 \cr
1 & 0 & 0 \cr
} \right)
\label{PSinA3}
\end{equation}
where $k$ and $l$ are arbitrary.

%**************************************
\section{Two-Weyl-graviton theory}
\label{section3}
%**************************************
\setcounter{equation}{0}
\setcounter{theorem}{0}
\setcounter{lemma}{0}
%~~~~~~~~~~~~~~~~

In this section we first show that the gauging of the extension $L' \equiv
so(4,2) \otimes \{e,n\}$
of the conformal algebra leads to the 2-Weyl-graviton
theory with action (\ref{strange0}). Then we construct the three-Weyl-graviton
theory.

\subsection{Curvatures}

We assign gauge fields to each of the 
generators of $L'$ and form a "big" gauge field
\bqn
h_{\mu}&=&h^A_{\mu}(T_A \otimes e) + h^{nA}_{~~\mu}(T_A \otimes n)=
\nonumber \\
       &=&h^A_{\mu}T_A + h^{nA}_{~~\mu}T^n_A.
\eqn 
The components are explicitly denoted
\be
e^a_{~\m}, \; \o_{\m}^{~ab}, \; b_{\m}, \; f^a_{~\m} , \;
e^{na}_{~~\m}, \; \o_{\;\m}^{n~ab}, \; b^n_{\m}, \; f^{na}_{~~\m} \, .
\label{bigfields}
\ee
The gauge parameters can be similarly expanded,
$\e=\e^AT_A+
\e^{nA}T^n_{A}$  and yield, when inserted into the
gauge transformation formula
$\d_{\e}h_{\m}=\pa_{\m}\e+[h_{\m},\e]\equiv \cd_{\m}\e$
\bqn
\d_{\e}h_{\m}^A&=&D^L_{\m} \e^{A},
\label{gaugevariafields} \\
\d_{\e}h^{nA}_{\m}&=& D^L_{\m} \e^{nA} 
+ f^{~~A}_{BC}
h^{nB}_{\m}\e^C.
\label{gaugevariafieldstar}
\eqn
where $D^L$ is the $L$-covariant derivative
\be
D^L \chi^{nA} = \pa_{\m}\chi^{nA}+f^{~~A}_{BC}h_{\m}^B\chi^{nC}
\ee
The second term in the right-hand side of (\ref{gaugevariafieldstar})
indicates that $h^{nA}_{\m}$ transform in the adjoint representation
of the original group.

The $L'$- curvatures are given by : 
\be
R_{\m\n}=R_{\m\n}^A T_A +R^{nA}_{\m\n}T^n_{A}= 2\pa_{[\m}h_{\n ]}+
[h_{\m},h_{\n}]
\ee 
(antisymmetrization is defined with the factor $(1/2)$ so that 
it is a projection operator).  Explicitly, one gets 
\be
R_{\m\n}^A=2\pa_{[\m}h^A_{\n ]}+f_{BC}^{~~A}h_{\m}^Bh_{\n}^C,
\label{courbures}
\ee
for the components along $L\otimes \{e\}$ and 
\be
R^{nA}_{\m\n}= 2D^L_{[\m}h_{~~\n ]}^{nA}
\label{curvastar}
\ee
for the components along $L\otimes \{n\}$.
Note in particular that the components along $L\otimes \{e\}$
are unchanged.
With the help of the formula 
$\d_{\e}R_{\m\n}=[R_{\m\n},\e]$
for the variation of the curvatures, we find
\begin{eqnarray}
\d_{\e}R_{\m\n}^A&=&f^{~~A}_{BC}R_{\m\n}^B\e^C
\label{gaugevariacurv}
\\
\d_{\e}R^{nA}_{\m\n}&=&f^{~~A}_{BC}(R_{\m\n}^B\e^{nC}+R^{nB}_{\m\n}\e^C).
\label{rotcurvstar}
\end{eqnarray}

Finally, the symmetric fields of the gravitons are obtained from
the "tetrads" $e^a_{~\m}$ and $e^{na}_{~~\m}$  through
\be
g_{\mu \nu}  = e^a_{~\m}e_{a \n}
\label{metricmetric}
\ee
and
\be
h_{\m\n}^n= 2 e^n_{(\m\n)} = e^a_{~\m}e^n_{a \n}+e^a_{~\n}e^n_{a \m}.
\label{perturbation}
\ee        
The tetrads $e^a_{~\m}$ are assumed to be invertible.  
World indices are lowered and raised with the metric $g_{\m\n}$
and its inverse $g^{\m\n}$.

\subsection{Action and constraints}
We take the action to be the natural extension
of the 1-graviton action of \cite{KTvN1}, namely,
the action is quadratic in the curvatures with
bilinear form given by (\ref{scalprodexten2}).  Taking $k=0$
for simplicity (see end of subsection {\bf \ref{sub33}}
for the general case), this leads to
\be
\stackrel{(2)}{S}[e,e^n,\o,\o^n,b, b^n, f,f^n]=2\int d^4x \ve^{\m\n\r\s}
\ve_{abcd}R_{\m\n}^{~~ ab}(M)
R_{\r\s}^{n~~ cd}(M^n).
\label{twoweylgravaction}
\ee
We also impose the constraints that the translation curvatures
be zero 
\be
\left\{
\begin{tabular}{ll}
$R_{\m\n}^a(P)=0$ \\
$R_{\m\n}^{na}(P^n)=0$ 
\end{tabular}
\right.
\label{constRP}
\ee
since, as we shall see in a moment, this is necessary for
invariance under conformal boosts.
The first constraint, being the same as the translation constraint
of the 1-graviton theory, can be solved in the
same way for $\o_{\m}^{~ab}$ to yield
\be
\o_{\m}^{~ab}=-2 g^{\rho \n}
e^{[a}_{~\rho} \pa_{[\m}e^{b]}_{~\n ]}+e^{a\r}e^{b\s}e_{c\m}
\pa_{[\r}e^{c}_{~\s ]}+2b^{[a}e^{b]}_{~\m}, \; \; \; b^a \equiv e^{a\m}b_\m.
\label{omega}
\ee  
Similarly, the second constraint can be solved for $\o_{\m}^{n~ab}$
and gives
\be
\o_{\m}^{n~ab}=D^{[a}e_{\;\m}^{n~b]}-D_{\m}e^{n[ba]}
+D^{[a}e_{~\m}^{n~b]}
+2b^{[a}e_{~\;\;\m}^{nb]}-2e^{[a}_{~\;\m}b^{nb]}
+2e^{[a}_{~\;\m}e^{nb]}_cb^c.
\label{omega*}
\ee                        
In the action (\ref{twoweylgravaction}), the $\omega$'s are not varied
independently, but must be regarded as functions of the other fields,
given by (\ref{omega}) and (\ref{omega*}).

The constraints (\ref{constRP}) are preserved under 
$M$-, $K$- and $D$-gauge transformations.  It follows 
that the $\omega$'s transform exactly as connections
under these
transformations, which implies,
in turn, that the curvatures transform as in (\ref{gaugevariacurv}) 
and (\ref{rotcurvstar}) (under those
same gauge transformations).  Therefore, the action (\ref{twoweylgravaction})
is invariant under the transformations generated
by $M_{ab} \otimes e$, $M_{ab} \otimes n$, $D \otimes e$, 
$D \otimes n$, $K_a \otimes e$ and $K_a \otimes n$ (using Lorentz
invariance of the quadratic form and the constraints).
The action is also trivially diffeomorphism invariant (because
it is a 4-form), i.e., invariant under
\be
\d_{\h_1} h^A_\m = {\cal L}_{\h_1} h^A_\m, \; \; \; \; 
\d_{\h_1} h^{nA}_\m = {\cal L}_{\h_1} h^{nA}_\m
\label{coord1}
\ee
as well as invariant under the transformations
\be 
\d_{\h_2} h^A_\m = 0 , \; \; \; \; \d_{\h_2} h^{nA}_\m
= {\cal L}_{\h_2} h^A_\m
\label{coord2}
\ee
under which only the components of the gauge field
along the generators $T_A \otimes n$
transform (one has $\d_{\h_2} R_{\m\n}^{~~ ab}(M) = 0$ and
$\d_{\h_2} R_{\m\n}^{n~~ ab}(M) = {\cal L}_{\h_2} R_{\m\n}^{~~ ab}(M)$).
Finally, the constraints are not preserved under $P$-gauge
transformations.  This means that the $\omega$'s acquire an 
extra variation term under translations. Following
the mechanism explained in
\cite{KTvN1,Mansouri:1977ej}, this has the net effect of trading
$P$-translations for the "coordinate"
transformations (\ref{coord1}) and (\ref{coord2}).

\subsection{Comparison with original action (\ref{strange0})}
\label{sub33}

To go from (\ref{twoweylgravaction}) to (\ref{strange0}), one
proceeds as follows:
\begin{enumerate}
\item  One first observes that $\o_{\m}^{ab}$ and $\o_{\m}^{n~ab}$
have been eliminated by means of the constraints;
\item One observes next that the fields $f^{a}_{~\m}$ and
$f^{na}_{~\m}$ taken together are auxiliary, 
in the sense that one can eliminate
them by using their own equations of motion.  More precisely,
the equation of motion for $f^{na}_{~\m}$ can be solved for 
$f^{a}_{~\m}$ and yield the same expression as in the 1-graviton
case, whereas the equation of motion for $f^{a}_{~\m}$ can be solved for
$f^{na}_{~\m}$.  One eliminates both $f^{a}_{~\m}$ and
$f^{na}_{~\m}$ using their equations of motion. 
One can in fact view these equations as other constraints involving the 
Ricci tensor \cite{PvN, TvN}.
[The explicit
on-shell expression of $f^{na}_{~\m}$ is not needed because
$f^{na}_{~\m}$ drops out once $f^{a}_{~\m}$ is on-shell.  The action
has schematically the form $a_{ij} (f^i - F^i) (f^{nj} -F^{nj})
+ S_0$ where $f^i$ (respectively $f^{ni}$) stands for
$f^{a}_{~\m}$ (respectively $f^{na}_{~\m}$) and where $F^i$, $F^{nj}$
and $S_0$ do not depend on the $f's$.  The
matrix $a_{ij}$ is invertible and the equations of motion for 
$f^i$ and $f^{ni}$ imply $f^i - F^i =0$ and $f^{nj} -F^{nj}$.
Once $f^i$ is put on-shell, $f^{ni}$ disappears.] 
\item Once this is done, one finds that the fields $b_\mu$
and $b^n_{\m}$ drop out from the action.  This is not surprising,
because these are the only fields left that transform under 
conformal boosts $K_a \otimes e$ and $K_a \otimes n$.
\item At this stage, the remaining fields in the action are
the tetrads $e^a_{~\m}$ and $e^{na}_{~\; \;\m}$.  They enter
only through the symmetric combinations (\ref{metricmetric}) and
(\ref{perturbation}) because of Lorentz invariance.  The resulting
action is in fact (\ref{strange0}) with gauge symmetries
(\ref{strange1}) and (\ref{strange2}).   This shows that indeed, the
action (\ref{strange0}) describes the gauging of the extension
$so(4,2) \otimes \{e,n\}$ of the conformal algebra.  Note that the
fields along the identity $e$ and the fields along the nilpotent
element $n$ play quite different roles.  The former are related
to Riemannian geometry through $g_{\m \n}$, which is a standard
Riemannian metric, while the latter appear in the end as fields 
propagating on that Riemannian spacetime.  
\end{enumerate} 

To summarize: one can construct an action with two Weyl fields $h^a_{\m \n}$
which are interacting, by gauging the extension (\ref{bigalgebra}) 
of the conformal group.  The action is (\ref{twoweylgravaction}).
There are two constraints $R_{\m\n}^{~~a}(P)=0$;  $R_{\m\n}^{n~~a}(P^n)=0$, 
and a pair of auxiliary fields $\left\{ f^a_{~\m},f^{na}_{~\;\;\m} \right\}$.
If we had taken the most general metric (\ref{PSinB}) in the commutative
algebra $\ca$ ($k \not= 0$), 
we would have obtained as action the sum of
the 1-graviton action  of \cite{KTvN1}  and of (\ref{twoweylgravaction}).  
The discussion
for this case proceeds straightforwardly, along exactly the same
lines as above and simply amounts to adding
to (\ref{strange0}) the Weyl action $\frac{1}{\a^2}\int 
\sqrt{-g} C^{\a \b \m \n}C_{\a \b \m \n}$ for the metric $g_{\m \n}$.

%~~~~~~~~~~~~~~~~~~~~~~~~~~~~~~~~~~~~~~~~~~~~~
\subsection{Three-graviton conformal gravity}
\label{subsection3gravitons}
%~~~~~~~~~~~~~~~~~~~~~~~~~~~~~~~~~~~~~~~~~~~~~

The $3$-graviton case - more generally, the $M$-graviton case - is
a direct extension of what we just found.
The underlying algebra to be gauged is in this case
the extension (\ref{2.20}) of
the conformal group, so
we have now three families of generators, fields, parameters and  curvatures
\be
\se{T_A}, \su{T_A},\sv{T_A},\se{h_{\m}^A}, \su{h_{\m}^A}, \sv{h_{\m}^A},
 \se{\e^A}, \su{\e^A}, \sv{\e^A},\se{R_{\m\n}^A}, \su{R_{\m\n}^A}, 
\sv{R_{\m\n}^A}.
\ee
The action reads
\be
I[h,\su{h}, \sv{h}]
=\int d^4x \ve^{\m\n\r\s}\ve_{abcd}[2{R_{\m\n}^{~~ab}(\se{M})}
{R_{\r\s}^{~~cd}(\sv{M})}+{R_{\m\n}^{~~ab}(\su{M})}{R_{\r\s}^{~~cd}(\su{M})}].
\ee
where we have set $k=l=0$ in the quadratic form (\ref{PSinA3}).
Keeping $k$ and $l \not=0$ would simply amount to adding to
the action given below the 1-graviton action of \cite{KTvN1} for
$\se{h_{\m}^A}$ and the 2-graviton action of the previous section
for $\se{h_{\m}^A}$ and $\su{h_{\m}^A}$.  The constraints are again
that all 
the components of the curvatures along the translation generators,
here $P_a \otimes e$, $P_a \otimes u$, $P_a \otimes u^2$, are equal to
zero.
One can solve these constraints for the $\omega$'s.
Furthermore, the $f$'s are again found to
be auxiliary fields.  Upon their elimination 
using their own equations 
of motion,
the action becomes a function of the symmetric combinations
\bqn
g_{\m\n}&=& \se{e}_{\;\m}^a \se{e}_{\;\n}^b\h_{ab}\\
h^u_{\m\n}&=& (\se{e}_{\;\m}^a  \su{e}_{\;\n}^b
+ \su{e}_{\;\m}^a \se{e}_{\;\n}^b) \h_{ab}\\
h^v_{\m\n}&=&  (\su{e}^a_{\m}\su{e}^b_{\n}+2\se{e}^a_{~(\m}\sv{e}^b_{\n)})
\h_{ab}
\eqn
of the tetrads only (the $b$'s drop out), invariant under
\bqn
\d^{{\rm{g.c.}}}_{\x^e,\x^u,\x^v}g_{\m\n}&=&
\cl_{\x^e}g_{\m\n},
\nonumber \\
\d^{{\rm{g.c.}}}_{\x^e,\x^u,\x^v}h^u_{\m\n}&=&\cl_{\x^e}h_{\m\n}^u
+\cl_{\x^u}g_{\m\n},
\nonumber \\
\d^{{\rm{g.c.}}}_{\x^e,\x^u,\x^v}h^v_{\m\n}&=&\cl_{\x^e}h_{\m\n}^v
+\cl_{\x^u}h_{\m\n}^u+\cl_{\x^v}g_{\m\n}.
\eqn
The dilatation part of the gauge transformations gives
\bqn
\d_{\f^e,\f^u,\f^v}g_{\m\n}&=&\f^eg_{\m\n},
\nonumber \\
\d_{\f^e,\f^u,\f^v}h^u_{\m\n}&=&\f^eh^u_{\m\n}+\f^ug_{\m\n},
\nonumber \\
\d_{\f^e,\f^u,\f^v}h^v_{\m\n}&=&\f^eh^v_{\m\n}+\f^uh^u_{\m\n}+\f^vg_{\m\n}.
\eqn        
We shall not write explicitly the final action but rather provide a method
for deriving it  -- as well as the $M$-graviton action -- 
directly from the 1-graviton action.  The approach explains
also the rationale behind the above construction and somehow "demystifies"
the occurrence of nilpotent elements in the underlying algebras.
%
%%%%%%%%%%%%%%%%%%%%%%%%%%%%%%
\section{Truncation of Taylor expansions}
\label{sectiontrivialisation}
%%%%%%%%%%%%%%%%%%%%%%%%%%%%%%
\setcounter{equation}{0}  
\setcounter{theorem}{0}  
\setcounter{lemma}{0}  
%~~~~~~~~~~~~~~~~~~~~~~~
%

\subsection{General theory}
The existence of the M-graviton action is in fact a consequence of the
following elementary observation.
Let $S^{(0)}[y^i]$ be an action invariant under the gauge transformations
\be
\d_{\e}y^i=R^i_{~\a}\e^{\a} 
\ee
(we adopt DeWitt's condensed notations).  The Noether identities read
\be
\frac{\d S^{(0)}}{\d y^i}R^i_{~\a}=0. 
\label{Noether}
\ee
By functional differentiation, one derives further identities
\be
\frac{\d^2 S^{(0)}}{\d y^j \d y^i} R^i_{~\a} + \frac{\d S^{(0)}}{\d y^i}
\frac{\d R^i_{~\a}}{\d y^j} = 0.
\label{Noetherderived}
\ee
It follows that the action
\be
S[y^i, Y^i] = k S^{(0)}[y^i] + S^{(1)}[y^i, Y^i], \; \; \; S^{(1)} \equiv
\frac{\d S^{(0)}}{\d y^i} Y^i
\label{actiontrunca1}
\ee
is invariant under the gauge transformations
\begin{eqnarray}
\d_{\e, \h}y^i&=&R^i_{~\a}\e^{\a}
\\
\d_{\e, \h} Y^i &=& \frac{\d R^i_{~\a}}{\d y^j} Y^j \e^{\a}
+R^i_{~\a} \h^{\a}.
\end{eqnarray}
If the field $y^i$ obeys the linear equation $D_{ij}y^j = 0$ in
the free limit ($S^{(0)Free}[y^i] \sim y^i D_{ij}y^j$ where $D_{ij}$ is a
symmetric differential operator), then the linearized equations of
motion following from (\ref{actiontrunca1}) will also be
$D_{ij}y^j = 0$, $D_{ij}Y^j = 0$.
One can view both the action (\ref{actiontrunca1}) and its gauge symmetries
as obtained from the original action (times $k$) by a limited Taylor
expansion 
\be
y^i \rightarrow y^i + x Y^i, \; \; \e^\a \rightarrow \e^\a + x\h^{\a}
\ee
in which one truncates to terms of degree $\leq 1$, which formally
amounts to assuming that $x$ is nilpotent of order $2$, $x^2 = 0$.
Invariance is automatic because it holds order by order in $x$.
The form  (\ref{actiontrunca1}) follows from a rescaling of $y^i$
in which one absorbs $k$ and $x$ (a similar rescaling must be
performed on $\h^{\a}$).
If one starts from the 1-Weyl-graviton action of \cite{KTvN1}
and performs this limited Taylor expansion, one gets the
2-Weyl-graviton action (\ref{twoweylgravaction}).

The 3-field action (respectively, M-field action) is obtained by
Taylor expanding and keeping  terms up to order 2 in $x$ (respectively, $M-1$)
\be
y^i \rightarrow y^i + x Y^i + x^2 Z^i, \; \; \e^\a \rightarrow \e^\a + x\h^{\a}
+ x^2 \l^{\a}.
\ee
For $M=3$, the action and gauge symmetries read
\be
S[y^i, Y^i, Z^i] = k S^{(0)}[y^i] + l S^{(1)}[y^i, Y^i] + S^{(2)}
[y^i, Y^i, Z^i]
\ee
with 
\be
S^{(2)} [y^i, Y^i, Z^i] = \frac{\d S^{(0)}}{\d y^i} Z^i +
\frac{1}{2}\frac{\d^2 S^{(0)}}{\d y^i \d y^j} Y^i Y^j
\ee
and
\begin{eqnarray}
\d_{\e, \h \lambda} y^i&=&R^i_{~\a}\e^{\a}
\\
\d_{\e, \h \lambda}Y^i &=& \frac{\d R^i_{~\a}}{\d y^j} Y^j \e^{\a}
+R^i_{~\a} \h^{\a}
\\
\d_{\e, \h \lambda}Z^i &=& (\frac{1}{2}\frac{\d^2 R^i_{~\a}}{\d y^k \d y^j} 
Y^k Y^j + \frac{\d R^i_{~\a}}{\d y^j} Z^j )\e^{\a}
\nonumber \\
&\;& +  \frac{\d R^i_{~\a}}{\d y^j} Y^j \h^{\a} + R^i_{~\a} \lambda^\a
\end{eqnarray}
Invariance of the action follows from the Noether identities (\ref{Noether})
and the
identities that follow by differentiating them once 
(see (\ref{Noetherderived})) and twice.

\subsection{Extension of Lie algebras}
The above formulas hold for any given theory.  If the original
theory is associated with Lie groups and gaugings, then the
multi-field theory has also an underlying Lie algebra structure,
which one easily identifies by expanding the fields and
the gauge transformations and truncating as above.  This
yields explicitly, for the $M$-field theory
\bqn
[\stackrel{(i)}{T_A},\stackrel{(j)}{T_B}]&=&f_{AB}^{~~C}\stackrel{(i+j)}{T_C}
 ~{\rm{for}} ~ 0\leq i+j<M
\nonumber \\
&=&0~{\rm{otherwise,}}
\eqn   

This is in fact $L \otimes \ca$ where $\ca$ is
the associative, commutative algebra generated by a 
unit and an element which is nilpotent of order $M$.
$L \otimes \ca$  is a truncation of the loop algebra
\be
[\stackrel{(i)}{T_A},\stackrel{(j)}{T_B}] = f_{AB}^{~~C}\stackrel{(i+j)}{T_C}
\; \; \; \; i,j \in Z
\ee
in which one retains only non-negative-moded elements  and drops
furthermore the $\stackrel{(i)}{T_A}$ with $i \geq M$.
Truncating is equivalent to assuming that the Taylor expansion parameter
fulfills $x^M= 0$, which clearly explains the emergence of nilpotent elements
in the associative algebra appearing in the tensor product $L \otimes \ca$. 

%%%%%%%%%%%%%%%%%%%%%%%%%%%%%%%
\section{Supersymmetrization}
%%%%%%%%%%%%%%%%%%%%%%%%%%%%%%%

After the discovery of ordinary supergravity
\cite{Freedman:1976xh, Deser:1976eh} based on the super
Poincar\'e algebra (which was later extended to the anti-de Sitter case
\cite{Freedman:1976aw,Fradkin:1976xz,Townsend:qa}),
a theory of supergravity was constructed, based on the conformal group
instead of the Poincar\'e group. Rather than using the Noether method,
the starting point was to consider curvatures of the form
$dh^A+f^A_{~BC}h^Ch^B$ where $f^A_{~BC}$ are the (constant) structure constants
of the $N=1$ superconformal algebra \cite{KTvN1}.
The notion that group theory can be used to construct gauge theories was used
to construct ordinary (not conformal) supergravity in several papers
both for the $N=1$ anti-de Sitter case
\cite{MacDowell:1977jt,Freedman:1976xh} and for the $N=2$ anti-de Sitter case
\cite{Chamseddine:1976bf}, \cite{Townsend:1977fz}.
As a warming-up exercise, first the bosonic case was
considered, the 15-parameter conformal group was gauged, and the action that
followed from this approach turned out to be the square of the Weyl tensor.
This was a bit of a surprise because one might have expected that also the
Maxwell action for the dilatation gauge field $b_{\m}$ (Weyl's invention)
should be present. The local conformal boost symmetry acted only on
$b_{\m}$ and this explained why $b_{\m}$ was absent from the action.

An extension of this approach to $N$-extended supergravities with
$U(N)$ internal symmetry group was begun in \cite{Ferrara:1977ij}.
To every generator of the superalgebra a gauge field $h^A_{\m}$ was
associated, but many of the $h^A_{\m}$ were nonpropagating. To eliminate
them, two approaches were followed : initially the equations of motion with
non-propagating fields were used to eliminate such auxiliary gauge fields,
but later constraints on the curvatures were used. The former method is
dynamical in origin, whereas the latter method is free from dynamics and
emphasizes the geometrical aspects. Clearly the latter method is more
fundamental.                                     

The complete theory of $N=1$ conformal supergravity was announced in
\cite{KTvN1}, and a detailed derivation was given in \cite{KTvN2}.
The relevant superalgebra is $su(2,2 \vert 1)$.
The gauge fields and corresponding generators are
$(e^a_{~\m}, P_a)$ for the translations,
$(\psi^{\a}_{~\m}, Q_{\a})$ for ordinary supersymmetry,
$(\o^{~ab}_{\m}, M_{ab})$ for the Lorentz transformations,
$(b_{\m}, D)$ for the dilatations, $(A_{\m}, A)$ for the $U(1)$ axial symmetry,
$(f^a_{~\m}, K_a)$ for the conformal boosts and
$(\varphi^{\a}_{~\m}, S_{\a})$ for conformal supersymmetry.
Of these fields, only $e^a_{~\m}$, $\psi^{\a}_{~\m}$ and $A_{\m}$ remain in the
final action. Due to local conformal boost gauge symmetry, the dilatation gauge
field $b_{\m}$ drops out of the action (just like for example the longitudinal
part of the electromagnetic field drops out of the Maxwell action). The spin
connection $\o^{~ab}_{\m}$ was expressed in terms of $e^a_{~\m}$,
$\psi^{\a}_{~\m}$, $b_{\m}$ by the constraint that the torsion
(the curvature of the translation generator $P_a$) vanishes, $R_{\m\n}^a(P)=0$.
The conformal gauge fields $f^a_{~\m}$ and $\varphi^{\a}_{~\m}$ were eliminated
as follows. The action was assumed to be quadratic in curvatures, i.e., of the
form $\e^{\m\n\r\s}R_{\m\n}^AR_{\r\s}^BQ_{AB}$, where $Q_{AB}$ were constant
Lorentz invariant tensors ($\e_{abcd}$, $\g_5$ and $\h^{ab}$). 
It preserved parity.
In addition there was a term $R_{\m\n}(A)R_{\r\s}(A)g^{\m\r}g^{\n\s}\sqrt{-g}$
which explicitly depended on the metric; the rest of the terms was affine.
Invariance of the action led to further constraints on the curvatures from
which $\varphi^{\a}_{~\m}$ could be eliminated in terms of other fields.
Finally, $f^a_{~\m}$ was eliminated from its own nonpropagating field
equations.

Later a new constraint, this time involving the Ricci tensor, was shown to
achieve the same result for the elimination of $f^a_{~\m}$  \cite{TvN}.
A final constraint on the supercurvatures was found in
\cite{PvN,vanNieuwenhuizen:cy} :
$R_{\m\n}(D)+\frac{i}{4}\e_{\m\n}^{~~\r\s}R_{\m\n}(A)=0$ was derived from
requiring closure of the gauge algebra. This constraint allowed one to
replace the term $R_{\m\n}(A)R_{\r\s}(A)g^{\m\r}g^{\n\s}\sqrt{-g}$
by $R_{\m\n}(D)R_{\r\s}(A)\e^{\m\n\r\s}$, so that now the whole action was
affine.
The final action read
\bqn
S=\int d^4x \e^{\m\n\r\s}[R_{\m\n}^{~~ab}(M)R_{\r\s}^{~~cd}(M)\e_{abcd}&+&
\a R_{\m\n}^{\a}(Q)(\g_5)_{\a\b}R_{\r\s}^{\b}(S)+
\nonumber \\
&+&\b  R_{\m\n}(A) R_{\r\s}(D)]
\eqn
where $\a$ and $\b$ are constants which were fixed by requiring invariance of
the action under all 24 local symmetries. The final set of constraints
 is given by 
\be
\left\{ \begin{tabular}{ll}
$R_{\m\n}^a(P)=0, ~~~~\g^{\m}R_{\m\n}(Q)=0$,
\\
$R_{\m\n}(M)+R_{\m\n}(D)+\bar{\psi}^{\l}R_{\n\l}(Q)=0$,
\\
$R_{\m\n}(D)+\frac{i}{4}\e_{\m\n}^{~~\r\s} R_{\r\s}(A)=0,$
\end{tabular}
\right.
\label{constraints}
\ee
where $R_{\m\n}(M)$ is the Ricci tensor.
The first three constraints are field equations in Poincar\'e theories,
but here they define the geometry. The last constraint is related to the
chirality-dualities which later became important.

Of course, conformal gravities and supergravities were also studied in other 
dimensions.
In $d=3$ for $N=0$ \cite{Deser:1981wh,Horne:1988jf},
$N=1$ \cite{Deser:sw,vanNieuwenhuizen:cx} and $N\geq 2$
\cite{Rocek:1985bk}. The actions were not squares of curvatures but rather
Chern-Simons actions
($d=3$ multi-graviton theories which break PT invariance have been considered 
recently in \cite{Boulanger:2000ni}).
Also in $d=5$ incomplete results were obtained
\cite{Ceresole:jp}. A review of conformal supergravities may be found 
in \cite{Fradkin:am}.
Our general analysis clearly shows how to construct
the theory of $M$ Weyl supermultiplets in interaction.  One
simply replaces the algebra $su(2,2 \vert 1)$ by its
extension $su(2,2 \vert 1) \otimes {\cal A}$ (where 
${\cal A}$ is generated by a nilpotent element of order $M$).

In the simplest case where ${\cal A}$ is spanned by a unity $e$ and a 
nilpotent element $n$ of order two with $e\star n=n$, $n\star n=0$ and with a 
non-positive-definite internal metric (\ref{PSinA}), the action for two
superconformal Weyl multiplets in interaction reads
\bqn
& S= \int d^4x \ve^{\m\n\r\s}
 [ 2R_{\m\n}^{~~ab}(M^e)\ve_{abcd} R_{\r\s}^{~~cd}(M^n)+ &
\nonumber \\
&+\a R_{\m\n}^{~~\a}(Q^e){(\g_5)}_{\a\b}R_{\r\s}^{~~\b}(S^n)
+\a R_{\m\n}^{n~\a}(Q^n){(\g_5)}_{\a\b}R_{\r\s}^{~\b}(S^n)+&
\nonumber \\
&+\b R_{\m\n}(A^e)R^n_{\r\s}(D^n)+\b R^n_{\m\n}({A}^n)R_{\r\s}(D^e)].&
\label{2superaction}
\eqn
The action (\ref{2superaction}) is defined on a set a constraints
which are, on the one hand, the constraints (\ref{constraints}) on
the generators $T_A\otimes e$ of $su(2,2 \vert 1) \otimes {\cal A}$, and
on the other hand, the constraints obtained from the set (\ref{constraints})
in a way similar to the way the action $S[y^i,Y^i]$
(\ref{actiontrunca1}) was obtained
from the action $S^{(0)}[y^i]$ in section {\bf{\ref{sectiontrivialisation}}}. 
That is, if we denote by $C^a[y^i]=0$, $a=1,...,K$ the set of constraints for 
the gauge 
theory of $su(2,2 \vert 1)$, the set of constraints for the gauge theory of 
$su(2,2 \vert 1) \otimes {\cal A}$ is given by the constraints   $C^a[y^i]=0$
together with the constraints  $D^a[y^i, Y^i]=0$,  
$D^a[y^i, Y^i]\equiv \frac{\d C^a}{\d y^i}Y^i$.
The fields $y^i$ are the gauge fields associated to the generators
$T_A\otimes e$, while the fields $Y^i$ are associated to the generators
$T_A\otimes n$.

\section{Conclusions}

In this paper, we have constructed interacting theories in $3+1$ dimensions,
involving a set of $M$ (super)-conformal Weyl multiplets.
These theories are obtained by replacing in a 
theory with one graviton all fields $y^i$ by $y^i+xY^i$ and all parameters 
$\e^{\a}$ by $\e^{\a}+x\h^{\a}$, and then truncating the action and 
transformation laws to terms at most linear in $x$. Because of this truncation,
these theories are truly interacting multi-graviton theories.
In principle our construction can be extended to lower dimensions and
higher dimensions, and to $\cal{N}$-extended conformal supergravities 
(which are Chern-Simons theories in odd dimensions).
These theories a priori suffer from the same physical drawbacks
as ordinary conformal (super)gravities since their linearized
versions have ghosts.  However, they exhibit interesting algebraic
structures that make them gauge-consistent (in the sense 
that the interacting theory has the same number of gauge symmetries
as the free theory).  This is a feature that could make them
of interest.  We have also explained why algebras with
nilpotent elements arise in this context, shedding thereby light on
previous work concerning the purely gravitational case.

The structures that we have encountered are reminiscent of loop extensions 
of $so(4,2)$.
However, if one performs the same analysis  for such extensions, one
loses the "triangular form" which
enabled one to explicitly eliminate the $\omega$'s and the $f$'s from
the action. For this reason, our construction does not appear to be 
immediately generalizable to that case.

\section*{Acknowledgements}

N.B. is supported by a grant from the Belgian 
{\it{Fonds pour la Formation \`a la Recherche dans l'Industrie et 
l'Agriculture }} (F.R.I.A.).
N.B. thanks Christiane Schomblond for helpul and kind discussions.
The work of N.B. and M.H. is supported in part by the "Actions
de Recherche Concert{\'e}es" of the "Direction de
la Recherche Scientifique - Communaut{\'e} Fran{\c c}aise
de Belgique", by a "P\^ole d'Attraction Interuniversitaire"
(Belgium), by IISN-Belgium (convention 4.4505.86),
by
Proyectos FONDECYT 1970151 and 7960001 (Chile) and
by the European Commission RTN programme HPRN-CT-00131,
in which they are associated to K. U. Leuven.


\begin{thebibliography}{99}
\bibitem{BDGH}
N.~Boulanger, T.~Damour, L.~Gualtieri and M. Henneaux,
Nucl.\ Phys.\ B {\bf 597}, 127 (2001)[hep-th/0007220];
%%CITATION = HEP-TH 0007220;%%
\\ 
{\it{Proceedings of International Conference on Quantization, Gauge Theory,
 and Strings: Conference Dedicated to the Memory of Professor Efim
Fradkin, Moscow, Russia, 5-10 Jun 2000}}, [hep-th/0009109]
%"No consistent cross-interactions for a collection ofmassless spin-2  fields"
%%CITATION = HEP-TH 0009109;%%"
\bibitem{Boulanger:2001he}
N.~Boulanger and M.~Henneaux,
%``A derivation of Weyl gravity,''
Annalen Phys.\  {\bf 10} (2001) 935
[hep-th/0106065]
%%CITATION = HEP-TH 0106065;%%
\bibitem{Barnich:vg}
G.~Barnich and M.~Henneaux,
%"Consistent Couplings Between Fields With A Gauge Freedom And Deformations 
%Of The Master Equation,"
Phys.\ Lett.\ B {\bf 311} (1993) 123
[hep-th/9304057]
%%CITATION = HEP-TH 9304057;%%
\bibitem{MH1} M. Henneaux,
%"Consistent interactions between gauge fields: The cohomological  approach",
Talk given at Conference on Secondary Calculus and Cohomological Physics,
Moscow, Russia, 24-31 Aug 1997, hep-th/9712226
%%CITATION = HEP-TH 9712226;%%"                 
\bibitem{KTvN1} 
M.~Kaku, P.~K.~Townsend and P.~van Nieuwenhuizen,
%"Gauge Theory Of The Conformal And Superconformal Group,"
Phys.\ Lett.\ B {\bf 69}, 304 (1977)
%%CITATION = PHLTA,B69,304;%%
\bibitem{Mansouri:pr} F. Mansouri,
%CONFORMAL GRAVITY AS A GAUGE THEORY.
Phys.\ Rev. \ Lett. \ {\bf{42}},\ 1021 (1979)
%%%CITATION = PRLTA,42,1021;%%
\bibitem{Mansouri:1977ej}
F.~Mansouri,
%"Superunified Theories Based On The Geometry Of Local (Super)Gauge 
%Invariance,"
Phys.\ Rev.\ D {\bf 16} (1977) 2456
%%CITATION = PHRVA,D16,2456;%%
\bibitem{PvN} P. van Nieuwenhuizen,
"Gauging of spacetime algebras", in "Quantum Groups and their applications
in Physics",{\it{ Proceedings of the 127th Enrico Fermi School at Varenna,
June 1996, p. 595}}   
\bibitem{KTvN2} M.~Kaku, P.~K.~Townsend and P.~van Nieuwenhuizen,
%"Properties Of Conformal Supergravity,"
Phys.\ Rev.\ D {\bf 17}, 3179 (1978)
%%CITATION = PHRVA,D17,3179;%%
\bibitem{TvN} P.~K.~Townsend and P.~van Nieuwenhuizen,
%"Simplifications Of Conformal Supergravity,"
Phys.\ Rev.\ D {\bf 19}, 3166 (1979)
%%CITATION = PHRVA,D19,3166;%%      
\bibitem{Lee:cp}
S.~C.~Lee and P.~van Nieuwenhuizen,
%"Counting Of States In Higher Derivative Field Theories,"
Phys.\ Rev.\ D {\bf 26} (1982) 934
%%CITATION = PHRVA,D26,934;%%
\bibitem{TseytlinSUGRA@25}
A.~A.~Tseytlin, Talk given at the conference "Supergravity at 25"
(CNYITP, Stony Brook, NY, December 1-2, 2001)
\bibitem{Hawking:2001yt}
S.~W.~Hawking and T.~Hertog,
%"Living with ghosts,"
[hep-th/0107088]
%%CITATION = HEP-TH 0107088;%%
\bibitem{Hull:1998vg}
C.~M.~Hull,
%"Timelike T-duality, de Sitter space, large N gauge theories and 
% topological field theory,"
JHEP {\bf 9807} (1998) 021
[hep-th/9806146]
%%CITATION = HEP-TH 9806146;%%}
\bibitem{Hull:2001ii}
C.~M.~Hull,
%"de Sitter space in supergravity and M theory,"
JHEP {\bf 0111} (2001) 012
[hep-th/0109213]
%%CITATION = HEP-TH 0109213;%%
\bibitem{Tseytlin:1995yw}
A.~A.~Tseytlin,
%"On gauge theories for nonsemisimple groups,"
Nucl.\ Phys.\ B {\bf 450} (1995) 231
[hep-th/9505129]
%%CITATION = HEP-TH 9505129;%%
\bibitem{MacDowell:1977jt}
S.~W.~MacDowell and F.~Mansouri,
%"Unified Geometric Theory Of Gravity And Supergravity,"
Phys.\ Rev.\ Lett.\  {\bf 38} (1977) 739
[Erratum-ibid.\  {\bf 38} (1977) 1376]
%%CITATION = PRLTA,38,739;%%
\bibitem{Wald1} C. Cutler and R. Wald,
%{\it A new type of gauge invariance for
%a collection of massless spin--$2$ fields/ I. Existence and uniqueness},
Class. Quant. Grav. {\bf 4} (1987) 1267
%%CITATION = CQGRD,4,1267;%%
\bibitem{ovrut} A. Hindawi, B. Ovrut, and D. Waldram,
%{\it Consistent
%spin--two coupling and quadratic gravitation},
Phys. Rev. {\bf D53}
(1996) 5583
\bibitem{Wald2} R. Wald,
%{\it A new type of gauge invariance for
%a collection of massless spin--$2$ fields/ II. Geometrical interpretation},
Class. Quant. Grav. {\bf 4} (1987) 1279
\bibitem{Anco:1998mf}
S.~C.~Anco,
%"Nonlinear gauge
%theories of a spin-two field and a spin-three-halves  field,"
Annals Phys.\  {\bf 270}, 52 (1998)
%%CITATION = APNYA,270,52;%%              
\bibitem{DV0} M.~Dubois-Violette, M.~Henneaux, M.~Talon and C.~M.~Viallet,
%"Some Results On Local Cohomologies In Field Theory,"
Phys.\ Lett.\ B {\bf 267}, 81 (1991)
%%CITATION = PHLTA,B267,81;%%
\bibitem{locality} M.~Henneaux,
%"Space-Time Locality Of The BRST Formalism,"
Commun.\ Math.\ Phys.\  {\bf 140}, 1 (1991)
%%CITATION = CMPHA,140,1;%%
\bibitem{DV} M.~Dubois-Violette, M.~Henneaux, M.~Talon and C.~M.~Viallet,
%"General solution of the consistency equation,"
Phys.\ Lett.\ B {\bf 289}, 361 (1992)
[hep-th/9206106]
%%CITATION = HEP-TH 9206106;%%
\bibitem{BBH1} G.~Barnich, F.~Brandt and M.~Henneaux,
%"Local Brst Cohomology In The Antifield Formalism. 1. General Theorems,"
Commun.\ Math.\ Phys.\  {\bf 174}, 57 (1995)
[hep-th/9405109]
%%CITATION = HEP-TH 9405109;%%
\bibitem{BBH2} G.~Barnich, F.~Brandt and M.~Henneaux,
%"Local Brst Cohomology In The Antifield Formalism. Ii. Application To 
%Yang-Mills Theory,"
Commun.\ Math.\ Phys.\  {\bf 174}, 93 (1995)
[hep-th/9405194]
%%CITATION = HEP-TH 9405194;%%
\bibitem{Freedman:1976xh}
D.~Z.~Freedman, P.~van Nieuwenhuizen and S.~Ferrara,
%"Progress Toward A Theory Of Supergravity,"
Phys.\ Rev.\ D {\bf 13} (1976) 3214
%%CITATION = PHRVA,D13,3214;%%
\bibitem{Deser:1976eh}
S.~Deser and B.~Zumino,
%"Consistent Supergravity,"
Phys.\ Lett.\ B {\bf 62} (1976) 335
%%CITATION = PHLTA,B62,335;%%
\bibitem{Freedman:1976aw}
D.~Z.~Freedman and A.~Das,
%"Gauge Internal Symmetry In Extended Supergravity,"
Nucl.\ Phys.\ B {\bf 120} (1977) 221
%%CITATION = NUPHA,B120,221;%%
\bibitem{Fradkin:1976xz}
E.~S.~Fradkin and M.~A.~Vasiliev,
%"Model Of Supergravity With Minimal Electromagnetic Interaction,"
LEBEDEV-76-197
\bibitem{Townsend:qa}
P.~K.~Townsend,
%"Cosmological Constant In Supergravity,"
Phys.\ Rev.\ D {\bf 15} (1977) 2802
%%CITATION = PHRVA,D15,2802;%%
\bibitem{Chamseddine:1976bf}
A.~H.~Chamseddine and P.~C.~West,
%"Supergravity As A Gauge Theory Of Supersymmetry,"
Nucl.\ Phys.\ B {\bf 129} (1977) 39
%%CITATION = NUPHA,B129,39;%%
\bibitem{Townsend:1977fz}
P.~K.~Townsend and P.~van Nieuwenhuizen,
%"Geometrical Interpretation Of Extended Supergravity,"
Phys.\ Lett.\ B {\bf 67} (1977) 439
%%CITATION = PHLTA,B67,439;%%
\bibitem{Ferrara:1977ij}
S.~Ferrara, M.~Kaku, P.~K.~Townsend and P.~van Nieuwenhuizen,
%"Gauging The Graded Conformal Group With Unitary Internal Symmetries,"
Nucl.\ Phys.\ B {\bf 129} (1977) 125
%%CITATION = NUPHA,B129,125;%%
\bibitem{vanNieuwenhuizen:cy}
P.~van Nieuwenhuizen,
%"Constraints In Conformal Simple Supergravity,"
ITP-SB-85-5
{\it  In *Gotsman, E. ( Ed.), Tauber, G. ( Ed.): From Su( 3) To Gravity*, 
369-382}
\bibitem{Deser:1981wh}
S.~Deser, R.~Jackiw and S.~Templeton,
%"Topologically Massive Gauge Theories,"
Annals Phys.\  {\bf 140} (1982) 372
[Erratum-ibid.\  {\bf 185} (1982) 406]
%%CITATION = APNYA,140,372;%%
\bibitem{Horne:1988jf}
J.~H.~Horne and E.~Witten,
%"Conformal Gravity In Three-Dimensions As A Gauge Theory,"
Phys.\ Rev.\ Lett.\  {\bf 62} (1989) 501
%%CITATION = PRLTA,62,501;%%
\bibitem{Deser:sw}
S.~Deser and J.~H.~Kay,
%``Topologically Massive Supergravity,''
Phys.\ Lett.\ B {\bf 120} (1983) 97
%%CITATION = PHLTA,B120,97;%%
\bibitem{vanNieuwenhuizen:cx}
P.~van Nieuwenhuizen,
%"D = 3 Conformal Supergravity And Chern-Simons Terms,"
Phys.\ Rev.\ D {\bf 32} (1985) 872
%%CITATION = PHRVA,D32,872;%%
\bibitem{Rocek:1985bk}
M.~Ro\v{c}ek and P.~van Nieuwenhuizen,
%"N >= 2 Supersymmetric Chern-Simons Terms As D = 3 Extended Conformal 
%Supergravity,"
Class.\ Quant.\ Grav.\  {\bf 3} (1986) 43
%%CITATION = CQGRD,3,43;%%
\bibitem{Boulanger:2000ni}
N.~Boulanger and L.~Gualtieri,
%"An exotic theory of massless spin-two fields in three dimensions,"
Class.\ Quant.\ Grav.\  {\bf 18} (2001) 1485
[hep-th/0012003]
%%CITATION = HEP-TH 0012003;%%
\bibitem{Ceresole:jp}
A.~Ceresole, A.~Lerda and P.~van Nieuwenhuizen,
%"On Five-Dimensional Gravitational Super Chern-Simons Terms,"
Phys.\ Rev.\ D {\bf 34} (1986) 1744
%%CITATION = PHRVA,D34,1744;%%
\bibitem{Fradkin:am}
E.~S.~Fradkin and A.~A.~Tseytlin,
%"Conformal Supergravity,"
Phys.\ Rept.\  {\bf 119} (1985) 233.
%%CITATION = PRPLC,119,233;%% }  
\end{thebibliography}
\end{document}